\documentclass{article}

\usepackage[final,nonatbib]{neurips_mlsb_2020}

\usepackage[utf8]{inputenc} 
\usepackage[T1]{fontenc}    

\usepackage{hyperref}       
\usepackage{url}            
\usepackage{booktabs}       
\usepackage{amsfonts}       
\usepackage{nicefrac}       
\usepackage{microtype}      
\usepackage{verbatim}
\usepackage{titlesec}

\usepackage{graphicx}
\usepackage{amssymb}

\usepackage{amsmath}
\usepackage{bm}
\usepackage{upgreek}

\usepackage{siunitx}
\usepackage[ruled,vlined]{algorithm2e}

\usepackage{subfigure}
\usepackage{caption}
\usepackage{tabularx}
\usepackage{floatrow}
\usepackage{multirow}
\usepackage{array}
\newfloatcommand{capbtabbox}{table}[][\FBwidth]

\makeatletter
\newcommand{\figcaption}[1]{\def\@captype{figure}\caption{#1}}
\newcommand{\tblcaption}[1]{\def\@captype{table}\caption{#1}}
\makeatother

\titlespacing{\section}{0pt}{*1.5}{*.5}
\titlespacing{\subsection}{0pt}{*1.5}{*.5}
\titlespacing{\subsubsection}{0pt}{*1.5}{*.5}
\titlespacing{\paragraph}{0pt}{*2}{*1}

\title{Generating 3D Molecular Structures Conditional on a Receptor Binding Site with Deep Generative Models}

\author{%
  Tomohide~Masuda\thanks{These authors contributed equally to this work} \\
  Comp. \& Systems Biology \\
  University of Pittsburgh \\
  Pittsburgh, PA 15213 \\
  \texttt{tmasuda@pitt.edu} \\
  \And
  Matthew~Ragoza\footnotemark[1] \\
  Comp. \& Systems Biology \\
  University of Pittsburgh \\
  Pittsburgh, PA 15213 \\
  \texttt{mtr22@pitt.edu} \\
  \And
  David Ryan Koes \\
  Comp. \& Systems Biology \\
  University of Pittsburgh \\
  Pittsburgh, PA 15213 \\
  \texttt{dkoes@pitt.edu} \\
}

\begin{document}

\maketitle

\begin{abstract}
   Deep generative models have been applied with increasing success to the generation of two dimensional molecules as SMILES strings and molecular graphs. In this work we describe for the first time a deep generative model that can generate three-dimensional (3D) molecular structures conditioned on a 3D binding site. Using convolutional neural networks, we encode atomic density grids into separate receptor and ligand latent spaces.  The ligand latent space is variational to support sampling of new molecules.  A decoder network generates atomic densities of novel ligands conditioned on the receptor.  Discrete atoms are then `fit' to these continuous densities to create molecular structures. 
   We show that valid and unique molecules can be readily sampled from the variational latent space defined by a reference `seed' structure and generated structures have reasonable interactions with the binding site.  As structures are sampled farther in latent space from the seed structure, the novelty of the generated structures increases, but the predicted binding affinity decreases. Overall, we demonstrate the feasibility of conditional 3D molecular structure generation and provide a starting point for methods that also explicitly optimize for desired molecular properties, such as high binding affinity.
  
\end{abstract}

\section{Introduction}

The principal goal of drug discovery and design is to find novel molecules that bind to and alter the activity of target proteins. This entails a long and costly pipeline in which a massive chemical space is repeatedly filtered down into the subset of compounds most likely to be active. Computational methods promise to enhance this process through rapid scoring and ranking of large libraries of molecules prior to hit confirmation with experimental assays \cite{Seifert2003}. This approach, called virtual screening, is now a well-established component of the modern drug development toolkit.

Despite its successful applications \cite{zhu2013hit,ripphausen2010quo}, virtual screening has inherent limitations. A complete exploration of chemical space is computationally intractable, and virtual screening does not propose compounds missing from existing libraries. Furthermore, it does not propose changes or alternatives to input compounds to improve their desired properties (e.g. lead optimization).

Generative models have the potential to overcome these limitations. Generative models try to learn the underlying distribution of chemical space and acquire the ability to sample this distribution to yield novel active compounds. Two of the most popular generative modeling methods are generative adversarial networks (GANs) \cite{goodfellow2014gan} and variational autoencoders (VAEs) \cite{kingma2013vae}.

Efforts thus far to develop generative models for computational drug design have mainly involved learning to map from two-dimensional molecules to a continuous latent space, then exploring that latent space to produce new molecules with desired properties. In this work, we describe for the first time a deep learning model that can represent and generate three-dimensional molecular structures conditional on the target 3D binding pocket. Using atomic density grids as input, we encode a 3D bound molecular structure, including the target binding pocket, to a latent space and decode it. We then apply a novel optimization algorithm to solve the inverse problem of fitting a 3D molecular structure to a generated atomic density grid and reliably produce valid molecules. We evaluate generated structures in terms of molecular validity, uniqueness and predicted binding affinity, as computed by a densely connected neural network implemented in gnina (\texttt{https://github.com/gnina}) \cite{Paul2020cross}. 

\begin{figure}[tb]
    \small
    \centering
    \centering
    \includegraphics[width=.925\linewidth]{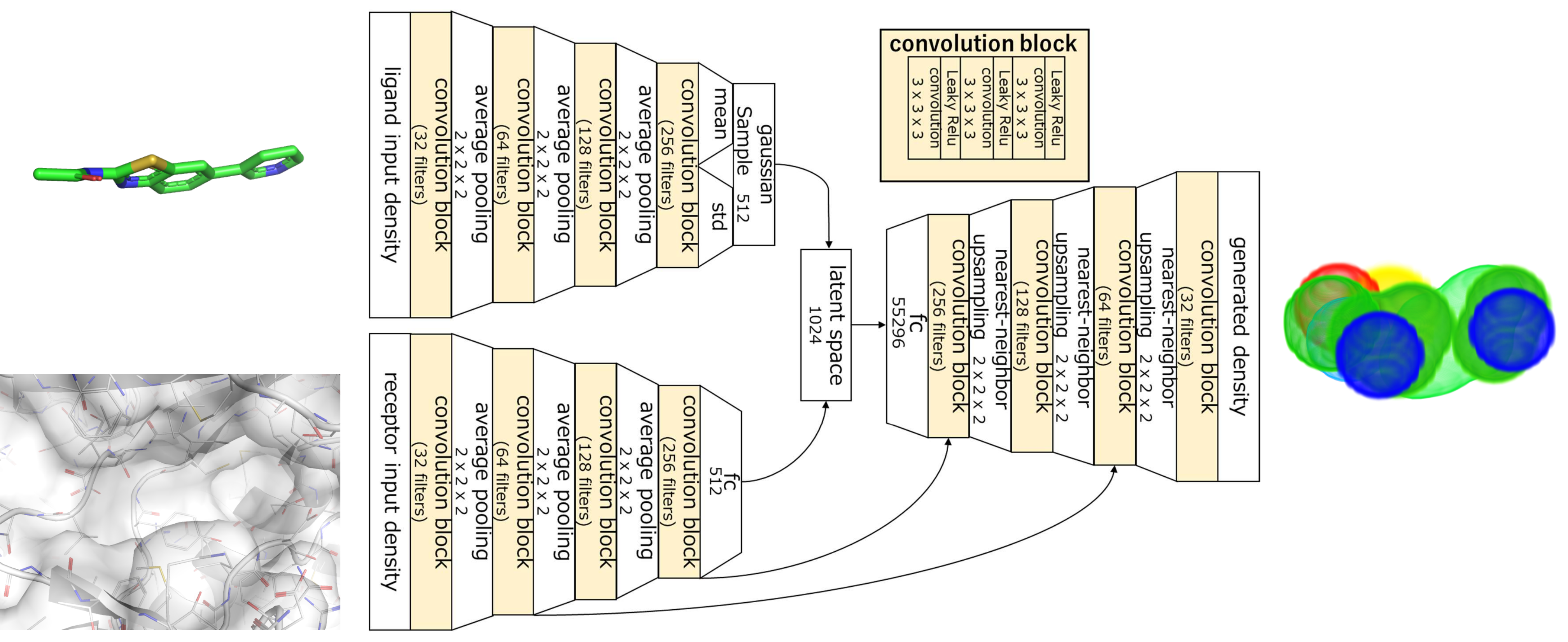}
    \caption{Model architecture. Encoder and decoder sub-networks are each a series of convolution block interleaved with either max-pooling (encoding) or nearest-neighbor upsampling (decoding). There are two skip connections from the receptor encoder to the decoder to propagate receptor structure information.
    }
    \label{fig:model}
\end{figure} 

\subsection{Related Work}

Given the advances made in computer vision over the past decade \cite{krizhevsky2012imagenet, lecun2015deep}, there has been a surge of interest in applying deep learning to drug discovery \cite{angermueller2016dl}. A close analogy to images for the 3D molecular structures is a grid of atomic densities, and this approach has been used successfully for binding discrimination for virtual screening \cite{ragoza2017cnn,imrie2018protein,wallach2015atomnet}, pose ranking for docking \cite{ragoza2017cnn, Paul2020cross}, and binding affinity prediction \cite{jimenez2018kdeep,Paul2020cross,stepniewska2018development,hassan2020rosenet}. 

In deep learning methods for generative modeling, generative adversarial networks (GANs) and latent variable models such as a variational autoencoders (VAEs)\cite{kingma2013vae} are widely used. Generative adversarial networks (GANs) \cite{goodfellow2014gan} are complementary to latent variable models \cite{larsen2016vaegan} and concurrently train a discriminative network to tell apart real and fake data, and a generative network to create samples that the discriminator classifies as real. Methods have been developed for generating image-like data from GANs \cite{radford2016dcgan} and for stabilizing their training dynamics \cite{gulrajani2017, arjovsky2017wgan}. In VAEs, an encoder and decoder network are trained to map between the input space and a set of latent random variables that follow a predetermined distribution allowing the model to sample from the prior and create novel data samples.

In the domain of deep generative models for molecular generation, recurrent neural networks (RNNs), which have been effective at modeling natural language, play a central role. To apply RNNs, the SMILES string format \cite{weininger1988smiles} is used as the molecular representation \cite{segler2017lm,Bjerrum2017}. However, RNNs based on SMILES have suffered from generating invalid SMILES strings that do not correspond to any molecules. Improvements have been made with grammar-based methods that ensure the correct context-free grammar of the output \cite{kusner2017gramvae, popova2018dl, Dai2018svae}. Additionally, similar molecules can have very different strings due to lack of permutation invariance, and randomization of SMILES traversal order was proposed to improve the quality of generated molecules \cite{Arus2019rsmiles}. VAEs in conjunction with SMILES were adopted to generate novel molecules from a trained continuous latent space \cite{gomez-bombarelli2016, Lim2018cvae}. Conditional VAEs, which can sample novel outputs conditioned on some relevant information, were also proposed for generating SMILES strings. These have been successfully applied to generating molecules that have desired molecular properties or related gene expressions\cite{Lim2018cvae, Mendez2020cgan}. 

Moreover, VAEs together with molecular graph representations \cite{simonovsky2018graphvae, jin2019jtvae}, which more naturally capture the invariances present in chemical structures, were proposed and can generate novel molecules step by step as molecular graph. Recently, VAEs, which take molecular graph representations and generate SELFIES \cite{Krenn2020selfies}, were also proposed \cite{Jacques2020optimol}.  
GANs have also been applied to molecular generative models and used with reinforcement learning to successfully generate novel molecules with desired properties \cite{Guimaraes2017, Sanchez2017}. In addition, GANs with low-dimensional mathematical representations of structural information of the protein-ligand complex were developed and used for pose scoring as part of the pose generation task \cite{Nguyen2020mathDL}.

Despite progress in generative modeling of molecules, strings and graphs are primarily two-dimensional approaches whereas atomic density grids \cite{ragoza2017cnn} allow learning from 3D conformations. Though grids are not rotation invariant, they are permutation invariant and have no discontinuities. Grids also have a fixed size, while SMILES strings and molecular graphs scale with the number of atoms and bonds. Atomic densities are continuous, differentiable, and highly GPU parallelizable as well \cite{sunseri2020libmolgrid}. Because they are continuous, density grids must be converted into a discrete representation to generate molecules. Previous efforts overcame this challenge by using a recurrent captioning network to output SMILES strings \cite{skalic2019shapevae}. However, this conversion relinquishes the three-dimensional nature of the generative model. To address this problem of generating 3D molecular structures, we introduce novel atom fitting and bond adding algorithms that can convert continuous grids to discrete molecular structures \cite{ragoza2020}. In this work we describe for the first time a deep generative model that can generate 3D molecular structures conditioned on a three-dimensional (3D) binding pocket using conditional VAEs \cite{Kingma2014cvae} in conjunction with GANs.

\section{Methods}

\subsection{Molecular grid representation}

We represent molecules in a grid format amenable to convolutional network training. To convert molecules to this format, we assign each atom a type based on a typing scheme that includes the element, aromaticity, hydrophobicity, and H-bond donor/acceptor state. Atoms are represented as continuous, Gaussian-like densities on a three-dimensional grid with separate channels for each atom type, analogous to an RGB image. Grids are centered on the molecule centroid and span a cube of side length 23.5{\AA} with 0.5{\AA} resolution, resulting in grid dimensions $N_x = N_y = N_z = 48$. To compensate for the lack of rotation invariance, molecules are randomly rotated during training. This is facilitated by computing grids on-the-fly using \texttt{libmolgrid}, an open-source, GPU-accelerated molecular gridding library \cite{sunseri2020libmolgrid}.

\subsection{Fitting molecules to generated grids}

While it is straightforward to compute the grid representation of a molecule, the inverse problem of converting a density grid into a discrete molecular structure does not have an analytic solution. Instead, we approach it as an optimization problem that we solve using an iterative algorithm that combines beam search and gradient descent (see \cite{ragoza2020} for more details) in order to identify a set of atoms the best fit a given atomic density grid.

The atom fitting algorithm returns a set of atom types and coordinates. To create a valid molecule, we then  assign bond based on distances between atoms, valence constraints, and atom type constraints. Since generated densities are often not ideal, we allow bond lengths as long as 4{\AA} if needed to generate a single molecule. Inaccuracies in the generated geometry are resolved through energy minimization as described below.

\subsection{Generative model architecture and training}

We trained our deep generative models as conditional variational autoencoders (CVAEs) \cite{Kingma2014cvae} in conjunction with a GAN loss using the atomic density grid format as their input and output. Generator networks were trained as deep convolutional CVAEs that took both a ligand and receptor density grid as input and produced a ligand density grid as output. The generator was trained to minimize the squared error, or L2 loss, of the output ligand density with respect to the input ligand density. An adversarial discriminative network was used to encourage the generated densities to match real densities from a more abstract perspective. The receptor binding site and ligand were encoded to distinct latent spaces by separate encoder branches, as seen in  Figure~\ref{fig:model}. The ligand latent space was variational, such that its latent vectors were sampled from a normal distribution that was parameterized by the ligand encoder (posterior). The ligand latent space was also constrained to follow a standard normal distribution overall by a KL divergence loss function (prior). The receptor latent space was deterministic and was used to condition the learned prior and posterior distributions. The encoded ligand and receptor latent vectors were concatenated and provided to the decoder, which produced a grid of ligand atomic density.

The encoder networks consisted of series of four convolution blocks with three convolution layers each having 32, 64, 128, or 256 filters. The encoder convolution blocks were interleaved with average pooling layers. The decoder network followed the same pattern in reverse, but interleaved its convolution blocks with nearest-neighbor upsampling instead. In two locations, convolutional outputs from the receptor encoder were concatenated with those of the ligand decoder in order to better propagate the 3D structural information of the receptor binding pocket at different resolutions while decoding the ligand.

The discriminator network had three convolution layers with filter sizes 16, 32, and 64, each followed by a leaky ReLU activation layer and an average pooling layer. The same kernel size of 3 was used in each of these convolution layers. The KL divergence, L2, and adversarial loss functions were combined with equal weight. Models were trained using the Adam optimization algorithm \cite{kingma2019adam} with hyperparameters $\alpha = 1e-5, \beta_1 = 0.9, \beta_2 = 0.999$ and a fixed learning rate policy for 1,000,000 iterations, using a batch size of 10. 

\subsection{Data sets}
In order to train deep neural networks to generate molecules in 3D, we utilized the CrossDocked2020 data set \cite{Paul2020cross}.  This set includes training and test splits that cluster similar protein targets together to avoid overlap between training and test sets.  It also attempts to mimic the drug discovery process by including ligand poses cross-docked against non-cognate receptor structure. Because our goal is to generate high-quality structures, we selected only low RMSD ($<2${\AA}) poses, yielding a total of 725,048 poses. We use a single training/test split. We note that although training data for 3D molecular structure generation is fundamentally limited to what is available in the Protein Data Bank (\texttt{https://www.rcsb.org}), the use of cross-docked structures provides a reasonable means of data augmentation, expanding the number of training structures by an order of magnitude.

\begin{figure}[tb]
    \small
    \centering
    \hfill
    \includegraphics[width=.45\linewidth]{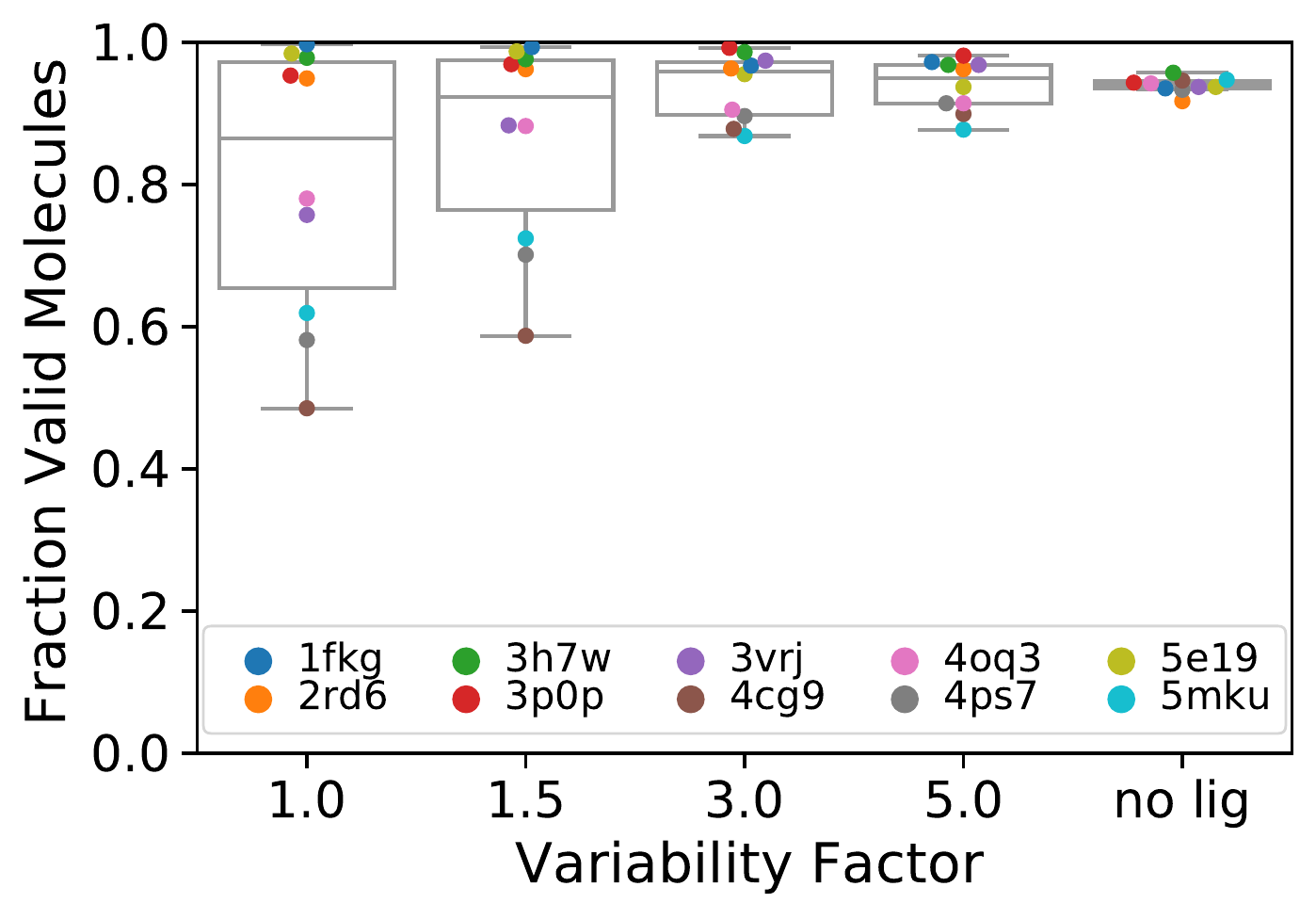}
    \hfill
    \includegraphics[width=.45\linewidth]{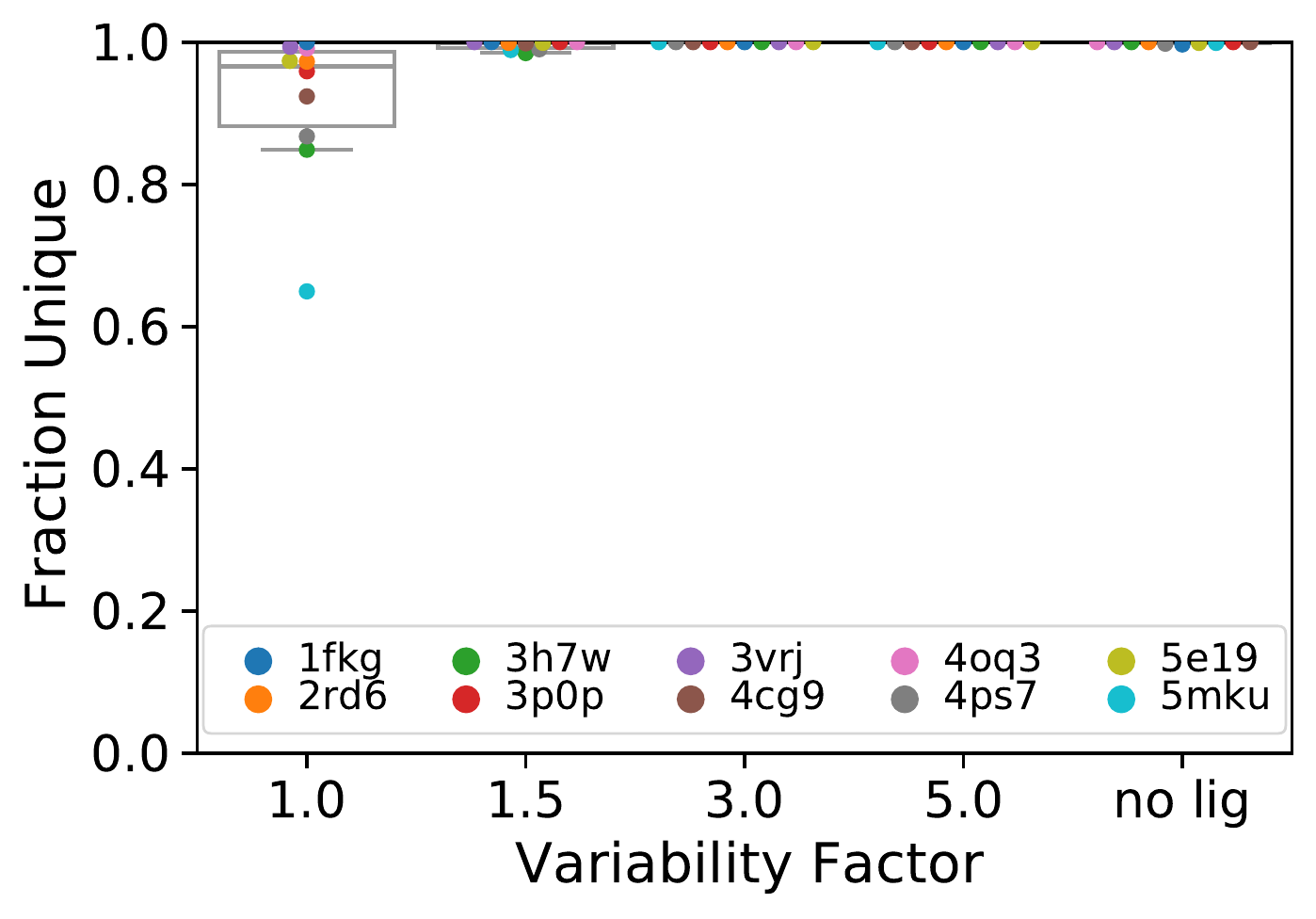}
    \hfill
    \caption{Ability to generate valid and unique molecules using ten complexes from the test set for different variability factors.}
    \label{fig:val_unq}
\end{figure}

\subsection{Evaluation procedure}
We evaluated structures generated by our trained models by providing an input seed molecule and receptor to the network (posterior sampling) and performing atom fitting and bond adding on the generated grids. We also evaluated prior (no lig) sampling where the ligand latent space is sampled as a standard normal distribution. In terms of drug discovery and development, posterior sampling corresponds to property optimization (e.g. hit to lead and lead optimization) and prior sampling to finding novel hit compounds.

First, we evaluated generated molecules in terms of their validity and uniqueness. A generated molecule is considered valid if does not contain disconnected fragments and RDKit is able to sanitize the molecule. Uniqueness was determined by equality comparison of canonical SMILES string representations of the valid molecules. 
We also examined the molecular similarity of generated molecules to the seed molecule using the MACCS fingerprint with the Tanimoto coefficient. 

Because generated molecules often had non-physical bond lengths and angles, we performed energy minimization for generated structure in the context of the fixed receptor using the UFF force field as implemented by RDKit.  The optimized molecule was then further optimized with respect to the receptor using the Vina scoring function as implemented by gnina.  Finally, we assessed the predicted binding affinity of optimized structures to the receptor using an ensemble of the densely connected CNN (\texttt{Dense}) models of gnina trained on CrossDocked2020 with five different seeds\cite{Paul2020cross}). For comparison, the reference seed ligand is also energy minimized with Vina. Vina optimization is necessary because the Dense CNN models are calibrated to such poses and because crystal ligands are not at an local optimum of the docking score, so comparisons to un-optimized structures would be misleading.
We report the root mean squared deviation (RMSD) between the initial generated structure and the final optimized structure. Larger RMSDs imply the generated molecule was geometrically or sterically problematic as the atoms had to be moved more to get to a local optimum.

We performed these generation and evaluation steps 1,000 times per each receptor-ligand complex evaluated. In each evaluation, different densities were generated due to the random sampling of the variational ligand latent space.
For this evaluation we sampled 10 receptor-ligand complexes from the test set (1fkg, 2rd6, 3h7w, 3p0p, 3vrj, 4cg9, 4oq3, 4ps7, 5e19 and 5mku). 
We also evaluated structure generation as we increased the variability factor for sampling the ligand latent space.  This is a multiplier of the generated variance \cite{skalic2019shapevae}.  A variability factor of 1.0 corresponds to the standard sampling used during training. Larger factors are expected to sample structures farther from the seed ligand.

\section{Results}
\subsection{Molecular validity, uniqueness and variability factor}

As shown in Figure~\ref{fig:val_unq}, we succeeded in producing valid molecules with an average rate of over 80\% at variability factor 1.0. More than 90\% of these molecules are chemically unique and, as the variability factor is increased, the percentage of unique molecules approaches 100\%.  This shows that even though the generated molecules are sampled conditioned on a seed molecule, they are not merely recapitulating the input. 

\begin{figure}[tb]
    \centering
    \includegraphics[width=\linewidth]{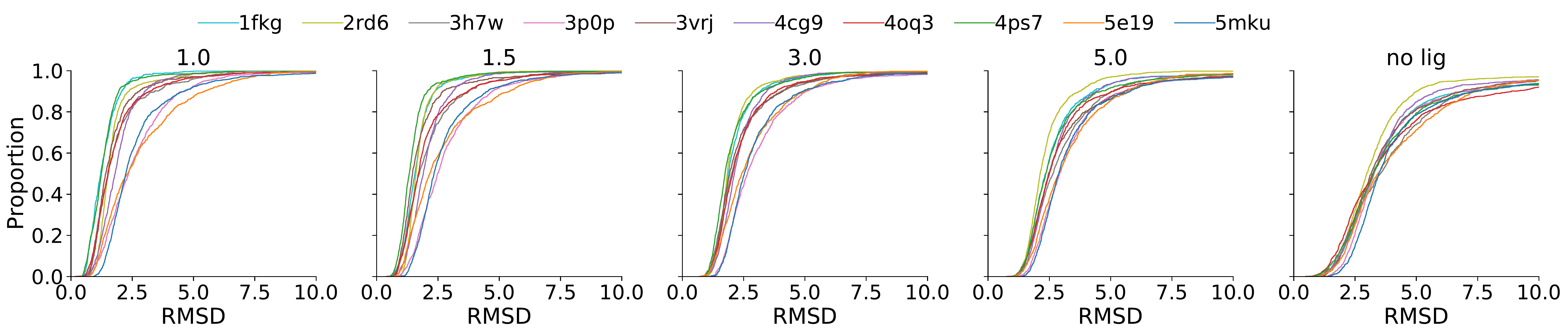}
        \includegraphics[width=\linewidth]{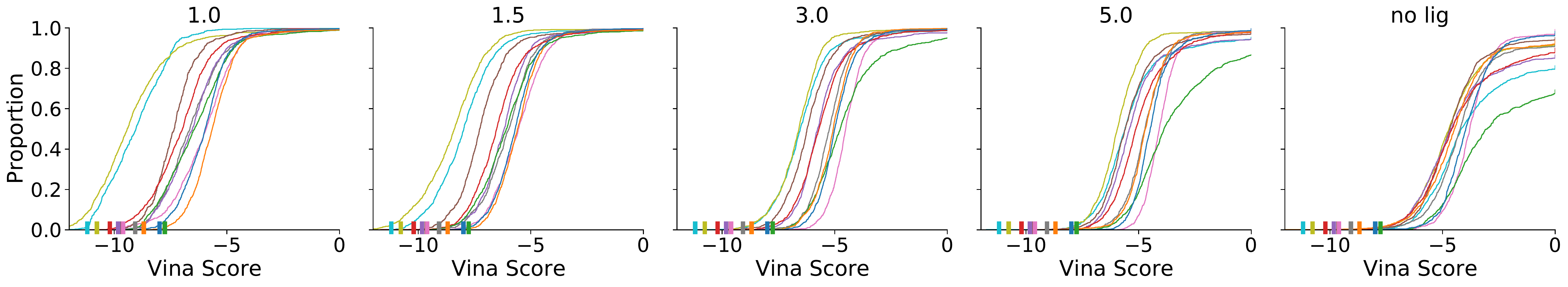}

    \caption{Cumulative distribution of RMSD change from the generated conformation to the UFF and Vina optimized compound (top) and the cumulative distribution of the Vina energy score (bottom).  Vina scores of the optimized reference ligand are shown as hashes on the axis.}
    \label{fig:rmsd_change}
\end{figure} 

\subsection{RMSD of optimization}

Figure~\ref{fig:rmsd_change} shows how significantly the energy minimization steps changed the molecules. RDKit minimization optimizes the pose as well as the internal bond lengths and angles in the context of the receptor binding site. Vina minimization only optimizes the pose orientation and internal rotatable bonds in the context of the receptor. As the variability factor is increased, generated molecules are more likely to be farther from the local minimum of the optimized molecule.  The overall median RMSD change is 1.65, 1.77, 2.09, 2.53, and 3.27 for variability factors of 1.0, 1.5, 3.0, 5.0 and prior sampling respectively.  An RMSD less than 2{\AA} is typically considered an acceptable amount of deviation for a docked pose \cite{Paul2020cross}. As the variability factor is increased, the generated structures deviate more from the seed molecule, resulting in higher RMSDs, but since they are also conditioned on the receptor, the RMSD change remains reasonable.

Figure~\ref{fig:rmsd_change} also shows cumulative distribution of Vina scores of the optimized molecules.  As the variability factor is increased, fewer good scoring (more negative) molecules are generated.  However, in all cases the majority of molecules have a negative score, indicating that reasonable optimized poses were identified, and, taking into consideration the RMSD results, these poses are not far, in most cases, from the unoptimized generated molecule.

\begin{figure}[tb]
    \centering
    \includegraphics[width=\linewidth]{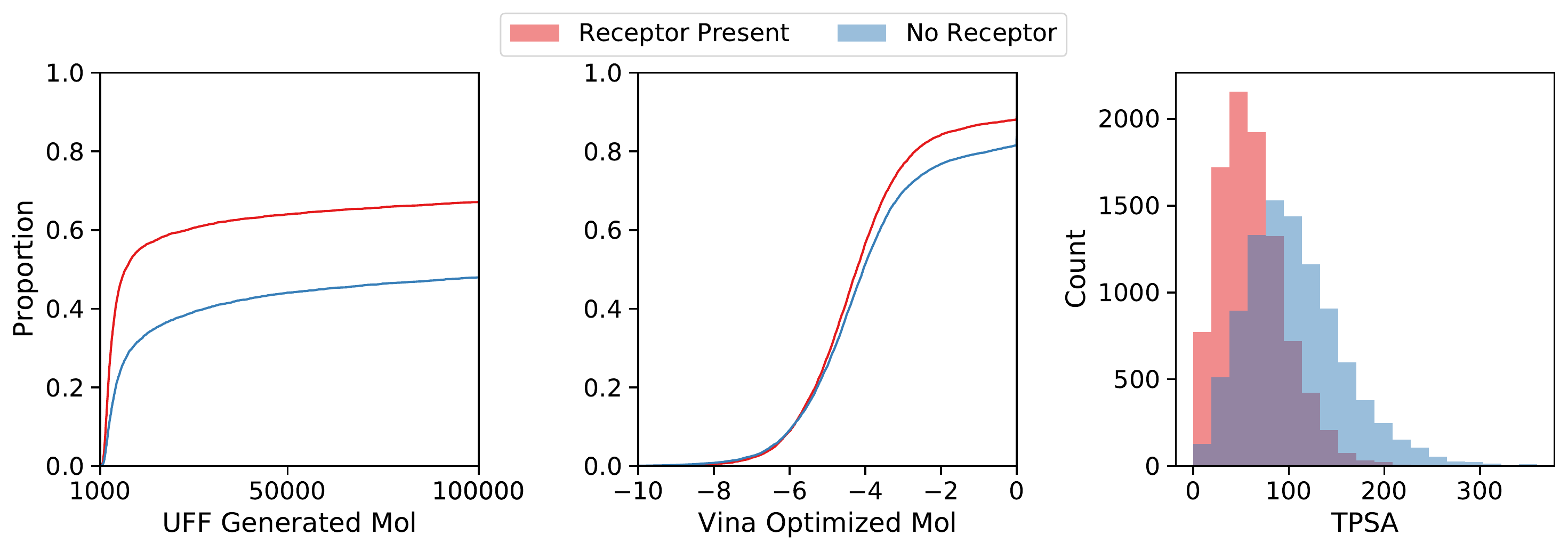}
    \caption{The cumulative distributions of UFF energy of the generated, unoptimized molecule and the Vina energy score of the optimized molecule for structures sampled from the prior (no lig) with and without a receptor structure.  Also shown (right) is the distribution of topological polar surface area (TPSA) of the generated molecules.}
    \label{fig:norec}
\end{figure}

\begin{figure}[ptb]
    \centering
    \includegraphics[width=\linewidth]{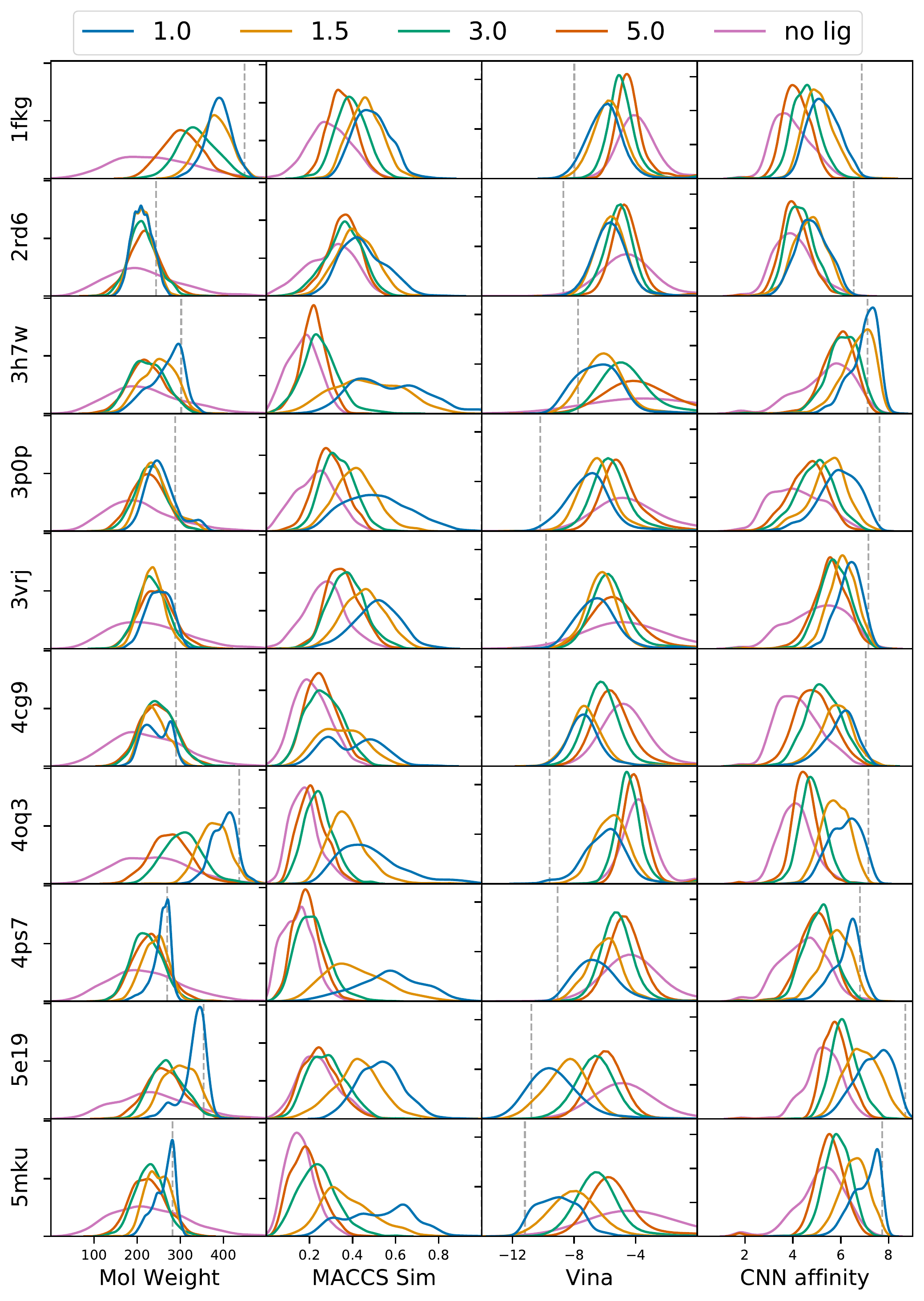}
    \caption{Distributions of molecular weight, similarity with the seed molecule, Vina energy score, and CNN affinity score for optimized generated molecules for different variability factors and complexes.
     Vertical lines show the corresponding value for the seed molecule.}
    \label{fig:mol_variability}
\end{figure}

\begin{figure}[tb]
    \centering
    \includegraphics[width=\linewidth]{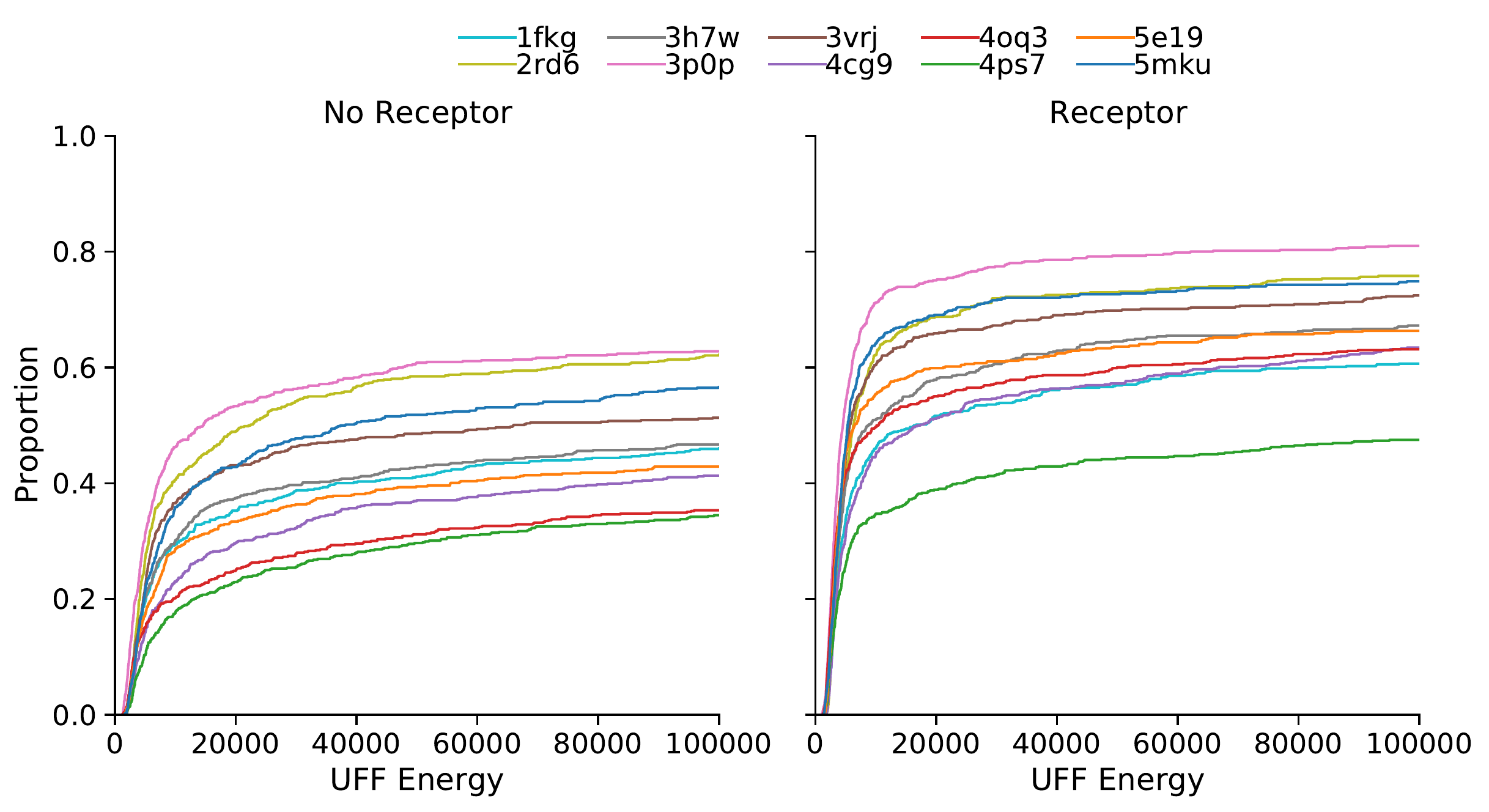}
    \caption{The cumulative distributions of UFF energy of the generated, unoptimized molecule for structures sampled from the prior (no lig) with and without a receptor structure broken out by individual PDBs.  Note that although the No Receptor structures are all drawn from the exact same distribution, the energy distribution will differ based on the receptor that is used when calculating the energy of the complex.}
    \label{fig:norecpdbs}
\end{figure}

\begin{figure}[tb]
    \centering
    \includegraphics[width=.38\linewidth]{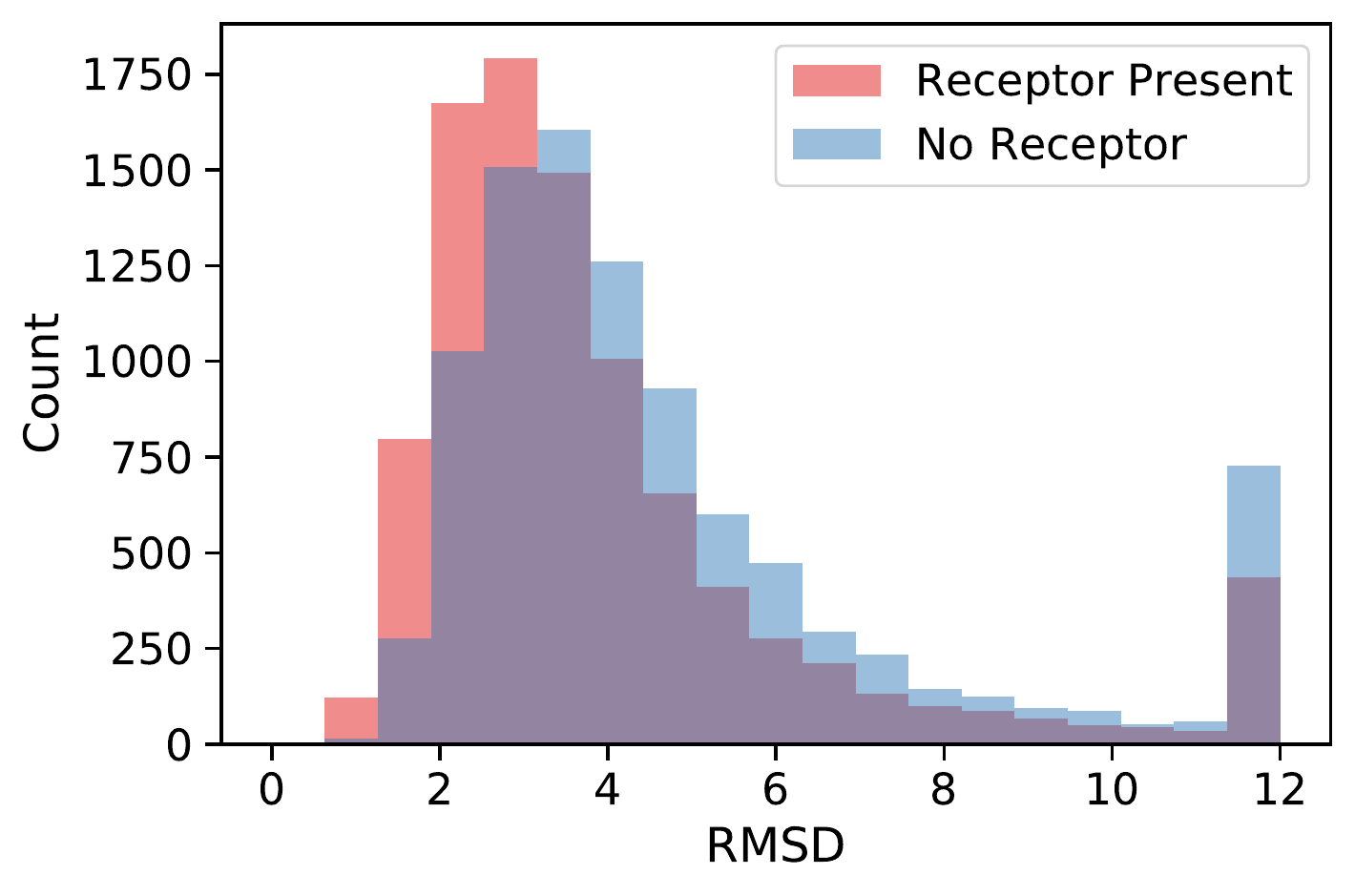}
    \hfill
    \includegraphics[width=.6\linewidth]{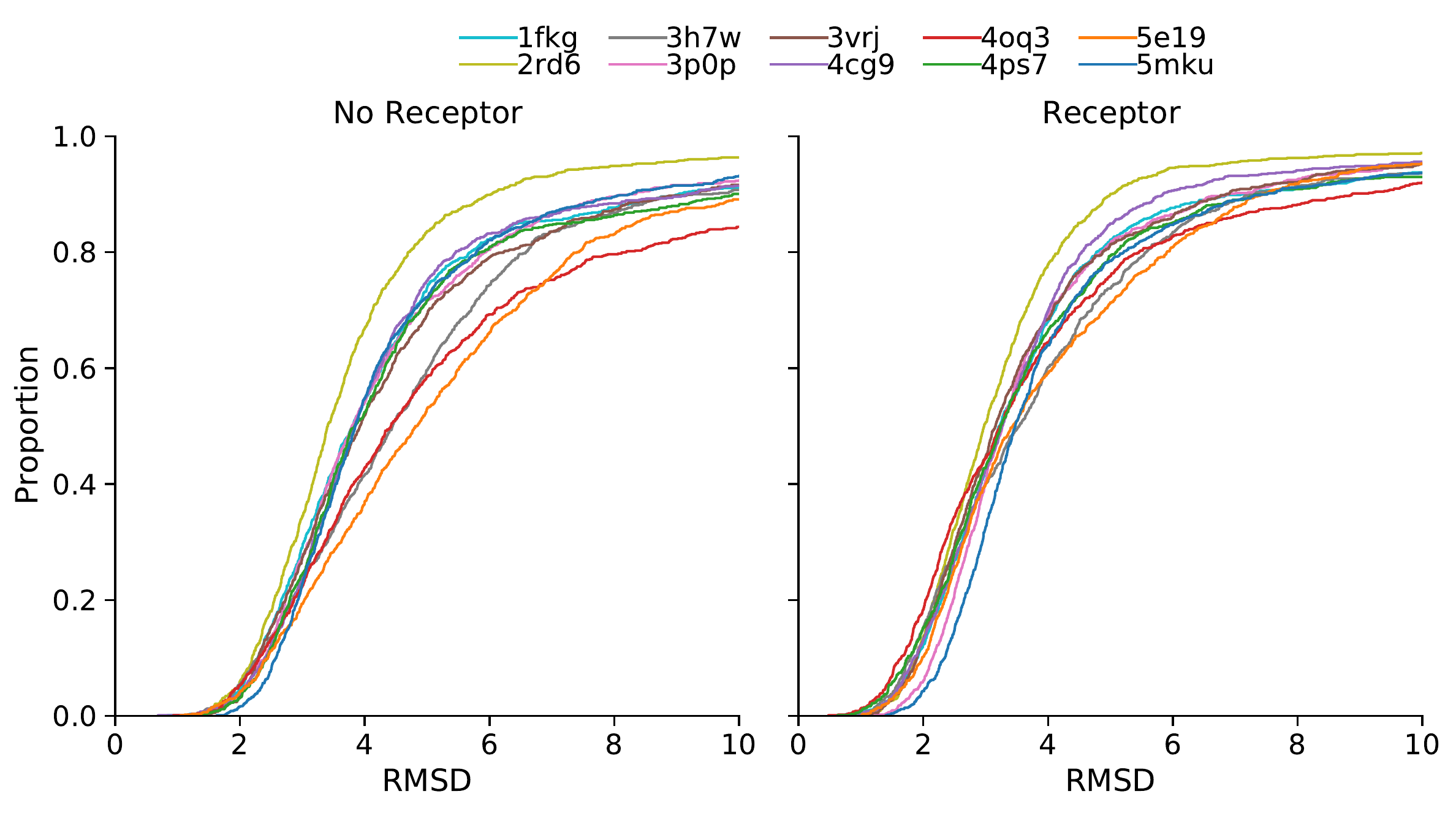}

    \caption{RMSD change after optimization of structures generated with and without a receptor when sampled from the prior (no lig).  (Left) Histogram of RMSDs.  Many more of the structures generated without the receptor require significant perturbations to converge to a local optimum. (Right) Cumulative distributions broken out by complex.}
    \label{fig:norecrmsd}
\end{figure}

\subsection{Molecular size, similarity, and affinity scores}

 Figure~\ref{fig:mol_variability} shows the change in the distribution of molecular size, similarity relative to the input molecule, and affinity scores as the variability factor is increased. In most  targets, the sizes of generated molecules were smaller than the seed molecule and larger variability factors made them smaller. Nonetheless, for most targets at least some molecules were generated that were larger than the seed.  In all cases, sampling without a seed ligand results in the widest distribution of generated molecule sizes.
 
 The affinity scores (Vina and CNN Affinity) and similarity also show a systematic tendency to shift away from the seed molecule as the variability factor is increased. Larger variability factors make affinity scores lower and similarity smaller (as expected). Every target had at least a few structures with better predicted affinity than the seed molecule when generated with variability factor 1.0. However, only two complexes, 3h7w and 5e19, had more than 15\% of their generated molecules have better Vina affinities (19\% and 17\% respectively), and these targets had 47\% and .6\% of generated molecules have a better CNN predicted affinity. 
 
\begin{figure}[tbhp]
    \centering
    \renewcommand{\tabcolsep}{1pt}
    \newcolumntype{Y}{>{\centering\arraybackslash}X}
    \begin{tabularx}{\textwidth}{p{1.25em}YY}
     & \textbf{3vrj} & \textbf{4ps7} \\
     \textbf{1.0} & \raisebox{-.5\height}{\includegraphics[width=\linewidth]{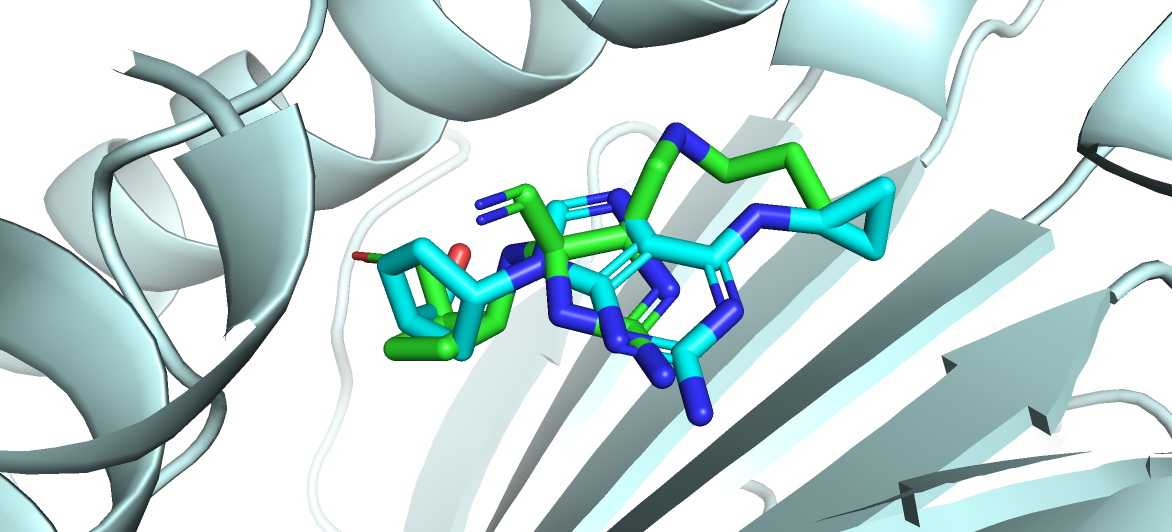}} 0.48 & \raisebox{-.5\height}{\includegraphics[width=\linewidth]{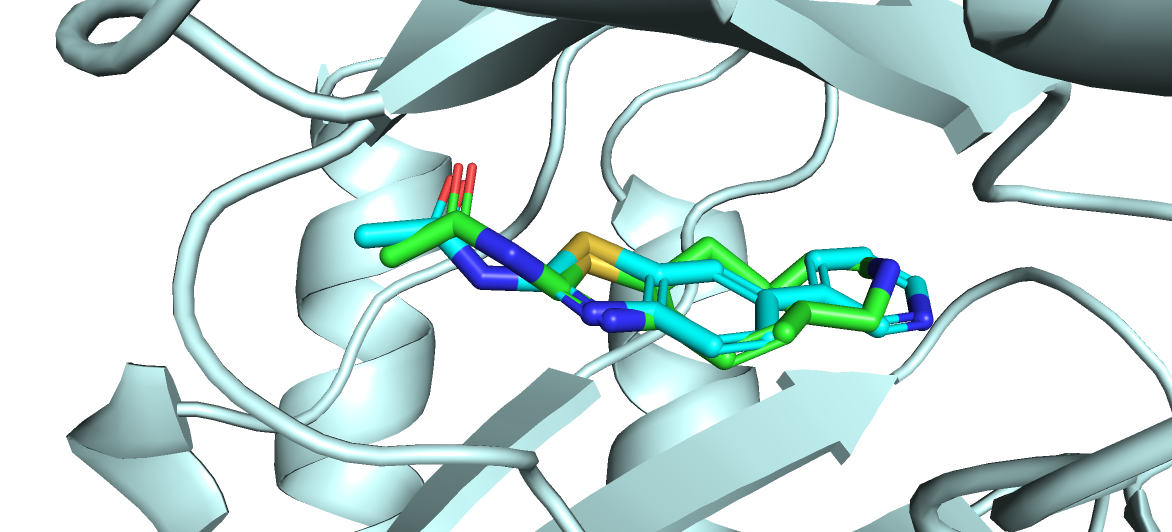}} 0.60 \\
     \textbf{1.5} & \raisebox{-.5\height}{\includegraphics[width=\linewidth]{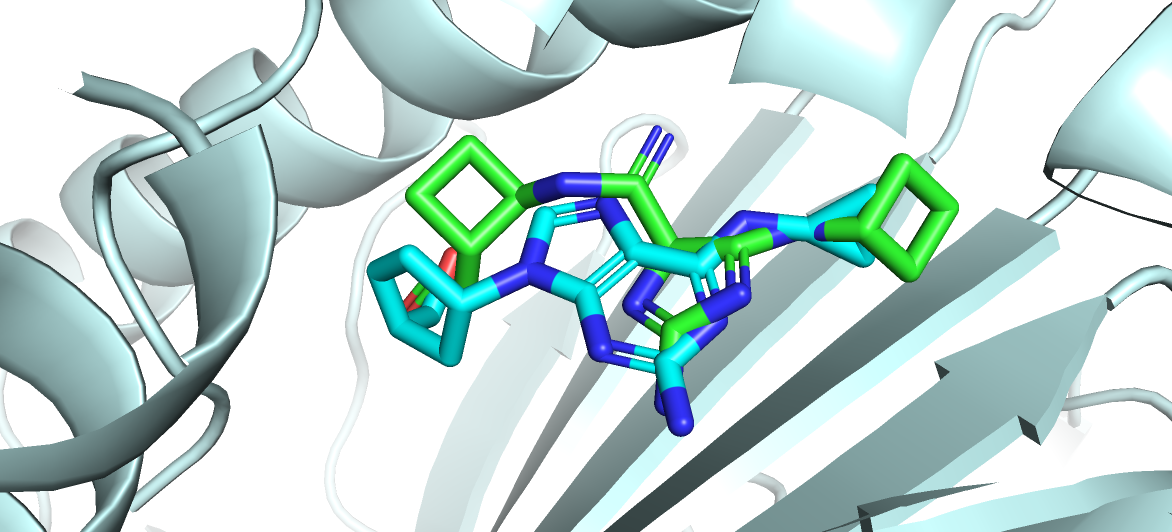}} 0.54 & \raisebox{-.5\height}{\includegraphics[width=\linewidth]{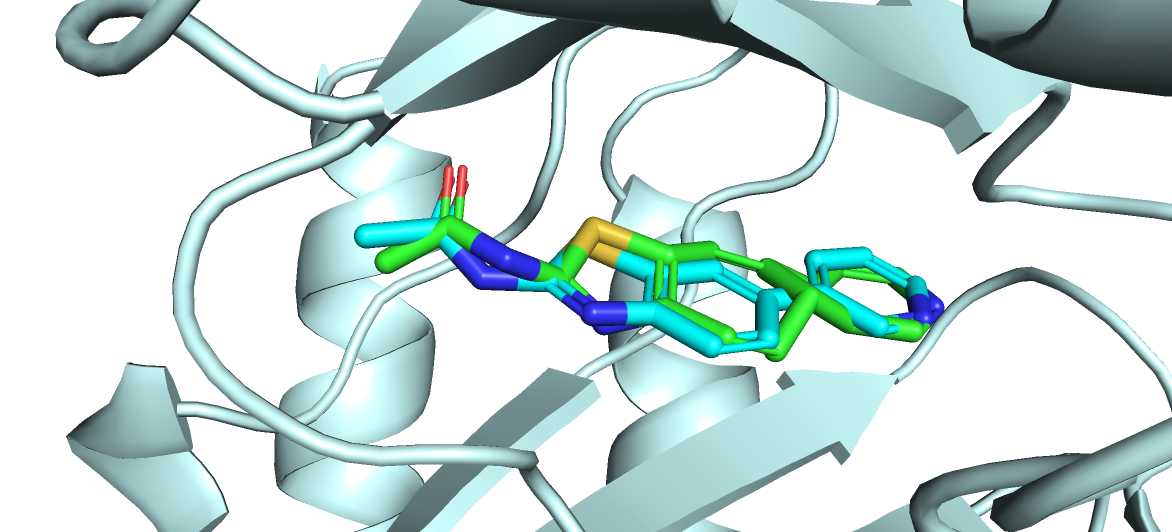}} 0.79 \\
     \textbf{3.0} & \raisebox{-.5\height}{\includegraphics[width=\linewidth]{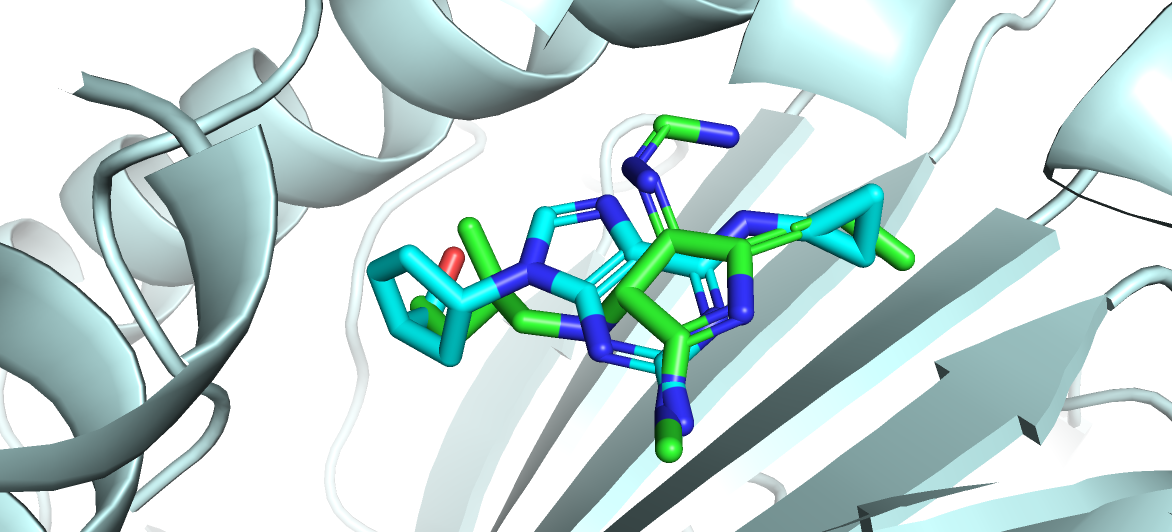}} 0.32 &  \raisebox{-.5\height}{\includegraphics[width=\linewidth]{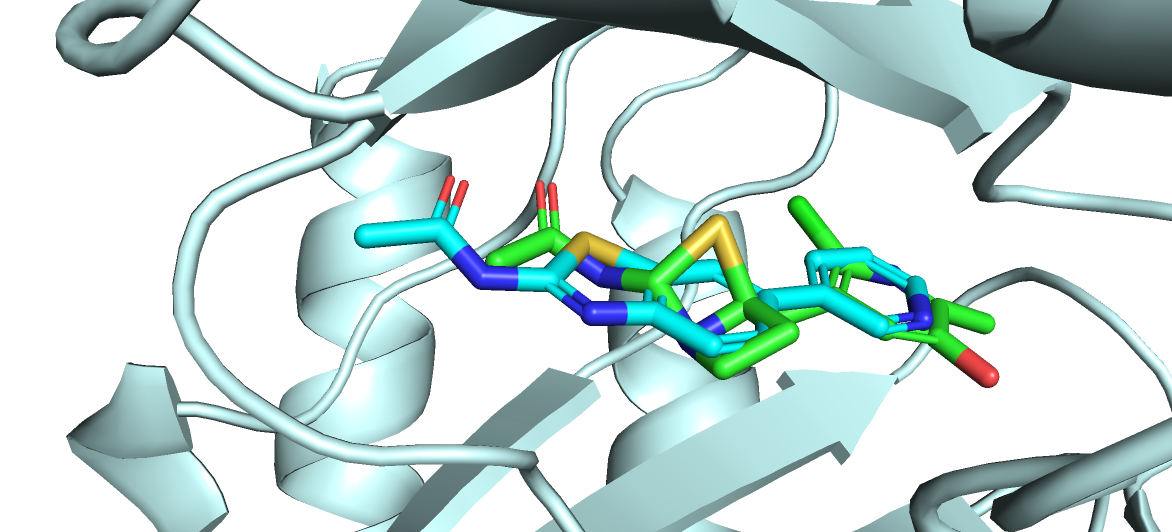}} 0.43\\
     \textbf{5.0} & \raisebox{-.5\height}{\includegraphics[width=\linewidth]{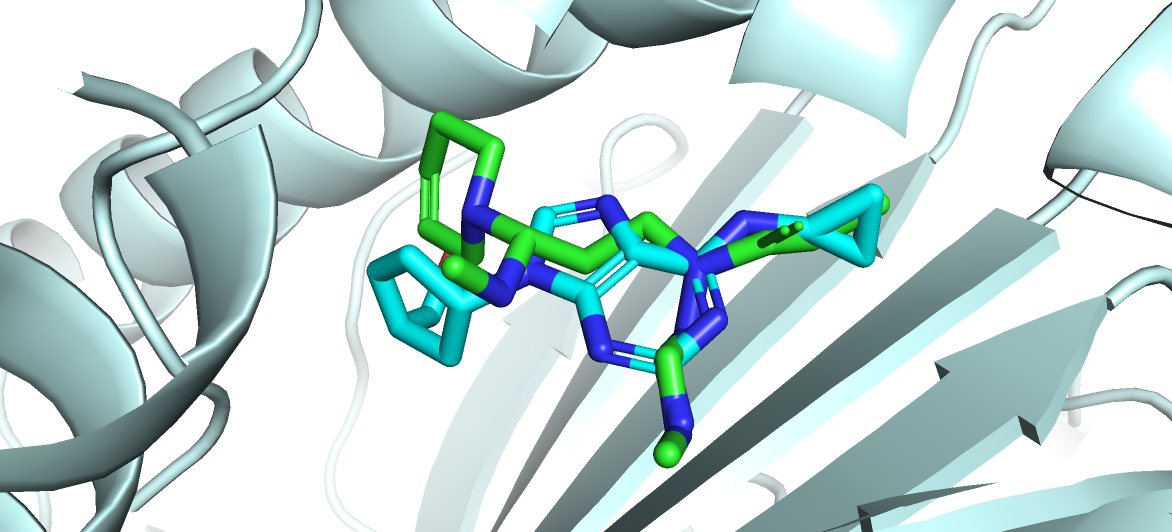}} 0.45 & \raisebox{-.5\height}{\includegraphics[width=\linewidth]{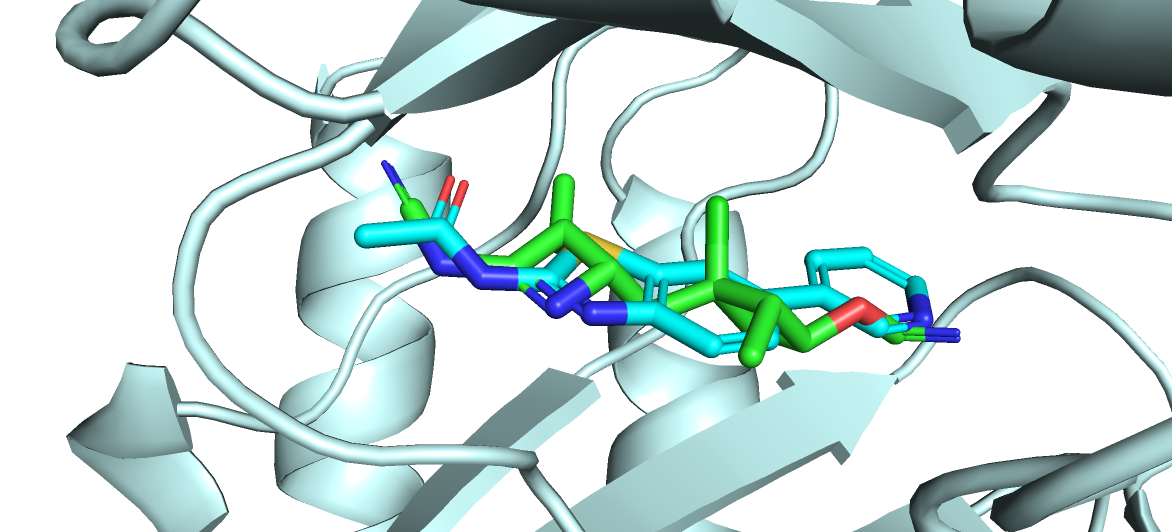}} 0.21\\ 
     \textbf{no lig} & \raisebox{-.5\height}{\includegraphics[width=\linewidth]{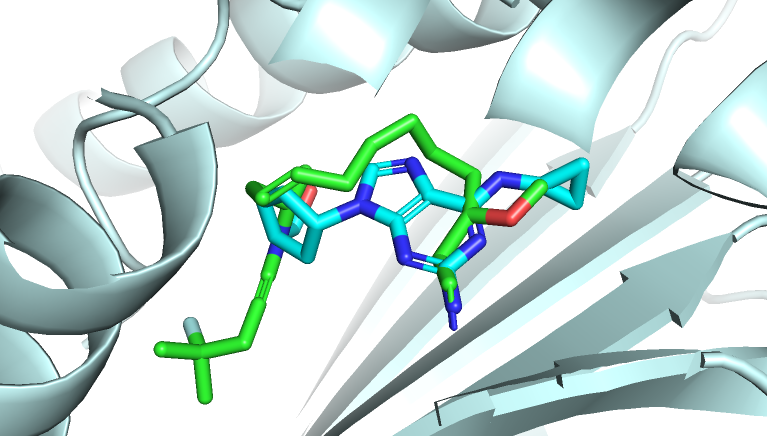}} 0.23 & \raisebox{-.5\height}{\includegraphics[width=\linewidth]{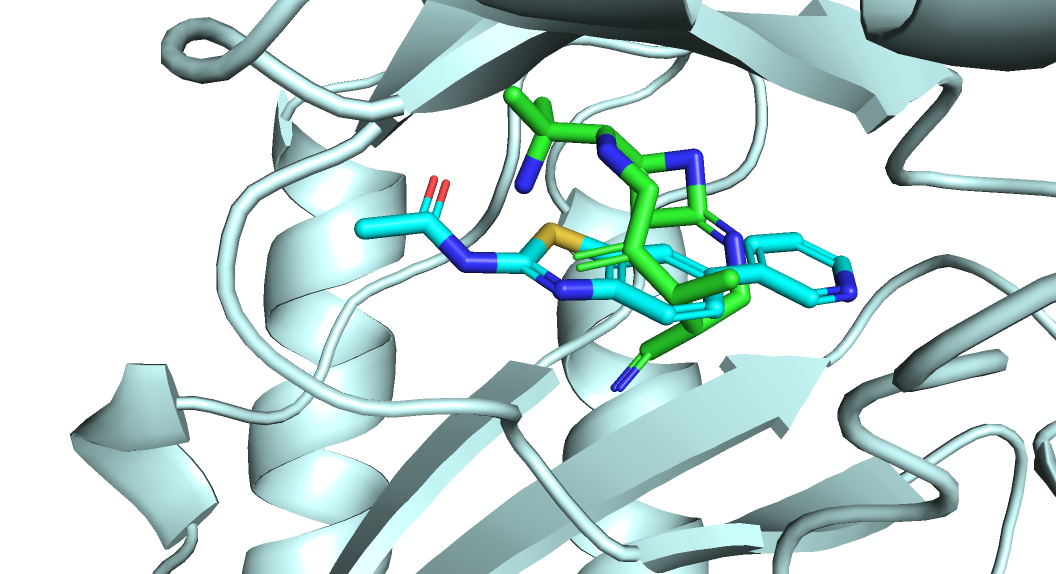}} 0.19\\      
    \end{tabularx}    
    \caption{Sampled binding poses for 3vrj and 4ps7 at different variability factors (rows). The seed molecule (blue) is shown with a generated pose (green) with its Tanimoto similarity coefficient.}
    \label{fig:binding}
\end{figure}

 As expected, as the variability factor is increased the similarity to the seed molecule decreases and, in fact, very few highly similar molecules are sampled. With variability factor 1.0, the percent of molecules with greater than 0.7 Tanimoto similarity ranged from 1\% to 22\%, but as the variability factor is increased, this percentage quickly goes to zero.
Although generated ligands have low chemical similarity to the seed ligand, as can be seen in Figure~\ref{fig:binding}, they appear to have high shape similarity.

\subsection{Relationship between affinity score and similarity}

As both predicted affinity and similarity decrease as the variability factor is increased, we investigate the relationship between these two values in Figure~\ref{fig:cnn_maccs}.  As expected, as the variability factor increases there is less similarity. Although in some cases these values are correlated (e.g. 4oq3 exhibits the highest Pearson correlation of 0.55 with a variability factor of 1.0), in many cases there is no correlation, and correlation is not necessary for high affinity ligands to be generated (e.g. 5e19 has a correlation of 0.04, but generates the best scoring ligands).  This suggests that success at generating high scoring structures is not solely due to generating chemically similar structures.

\subsection{Conditionality on receptor}
We investigate the extent that the generated structures are dependent on the receptor structure in Figure~\ref{fig:norec} which shows the overall cumulative distribution of energy scores of generated structures before and after optimization (distributions broken out by individual complexes are shown in Figure~\ref{fig:norecpdbs}).  Structures were generated with and without a receptor structure when sampling from the prior (no seed ligand provided).  As expected, structures generated without the receptor were much more likely to have a high (unfavorable) energy before optimization.  However, our optimization protocol reduces this difference significantly, although structures generated without the receptor have a higher RMSD after optimization (Figure~\ref{fig:norecrmsd}).  In addition to having distinctly different receptor interaction property distributions, the molecules have different molecular properties, most significantly the topological polar surface area, also shown in Figure~\ref{fig:norec}.  We observe that receptors with solvent inaccessible, hydrophobic binding sites (e.g. 3h7w and 5mku) produce molecules with a lower TPSA while solvent exposed sites (e.g. 1fkg and 4oq3) exhibit a higher TPSA.  Accordingly, molecules generated with no receptor also have significantly higher TPSA values.

\subsection{Example structures}
 High scoring generated structures for two targets sampled at each variability factor are displayed in Figure~\ref{fig:binding}. As the variability factor is increased, the generated molecule becomes visually distinct from the seed molecule. For 3vrj, the generated structures sampled with variability factors over 1.5 have a Tanimoto coefficient less than 0.5 relative to the seed molecule and don't retain the scaffold. In particular, the aromatic ring disappears and at a variability of 5.0 the generated structure is an open chain. For 4ps7, the structures generated with variability factors 1.0 and 1.5 look similar to the seed molecule and interactions such as hydrogen bonds are kept. Although their Tanimoto coefficients are less than 0.8, the scaffold of the seed molecule\cite{Collier20154ps7} is retained. The pyrydine substructure included in the seed molecule is replaced with a larger non-aromatic ring and modification of this heterocyclic structure contributes to the reduction of Tanimoto coefficient. In the case of variability factor 3.0, the generated molecule has a distorted structure around the scaffold of this seed molecule and this makes it difficult to retain hydrogen bonds with the receptor.  In the case of the factor 5.0, the generated molecule is totally different from the seed molecule and its Tanimoto coefficient is around 0.2. As expected, sampling without a seed molecule results in structures that are completely different both in terms of their chemical structure and their binding mode. Overall, molecules generated using larger variability factors tend to have unconventional scaffolds, pointing to the need to explicitly train models to generate drug-like molecules.

\section{Discussion}

By training deep generative models on atom density grids and applying our atom fitting and bond adding algorithms, we have developed the first deep learning model for generating fully three-dimensional molecular structures within binding pockets. 
Our model utilizes conditional VAEs and GANs and is conditional on a three-dimensional binding site. Whereas some works on molecular generative models using conditional VAEs have been developed, and they can successfully generate molecules, they are conditional on properties of the ligand, not the receptor \cite{Lim2018cvae}. Because our model generates fully three-dimensional molecular structures, the structure of the binding site is taken into account.  Unique to our approach, the importance of the receptor is emphasized in the network architecture through the propagation of receptor information at different spatial resolutions to the decoder.
 Our model was able to generate valid molecules at a rate greater than 80\% on average and most of the valid molecules were unique. When the ligand space is sampled close to a reference seed compound, it is possible to generate novel structures with better predicted binding affinities and similar size to the seed compound. Although generated structures typically do not produce ideal molecular geometries, usually only minimal changes ($<2${\AA} RMSD) are required to obtain a proper geometry.  As the variability of sampling is increased, larger changes in molecular structure are observed and fewer of these structures are close to locally optimal conformations.

Future work will seek to further regularize the ligand latent space to impose a drug-likeness bias as others have done \cite{Lim2018cvae}. We are also interested in exploring ways to generate structures with better predicted binding affinities, either through exploration of the ligand latent space or by training models that are explicitly conditioned on binding affinity or other molecular properties. Such approaches are likely necessary in order to enable the effective generation of structures from a receptor structure without a reference seed ligand.
Nonetheless, this work demonstrates the feasibility of conditional 3D molecular structure generation and provides a starting point for future improvements.


\section*{Acknowledgements}
 This work is supported by R01GM108340 from the National Institute of General Medical Sciences, is supported in part by the University of Pittsburgh Center for Research Computing through the resources provided, and used the Extreme Science and Engineering Discovery Environment (XSEDE), which is supported by National Science Foundation grant number ACI-1548562 through the Bridges GPU-AI resource allocation TG-MCB190049.

\small
\bibliography{biblio}

\newpage
\renewcommand\thefigure{S\arabic{figure}}    
\setcounter{figure}{0}  
\setcounter{page}{1}
\section*{Supplementary Materials}
\begin{figure}[h]
    \centering
    \includegraphics[width=\linewidth]{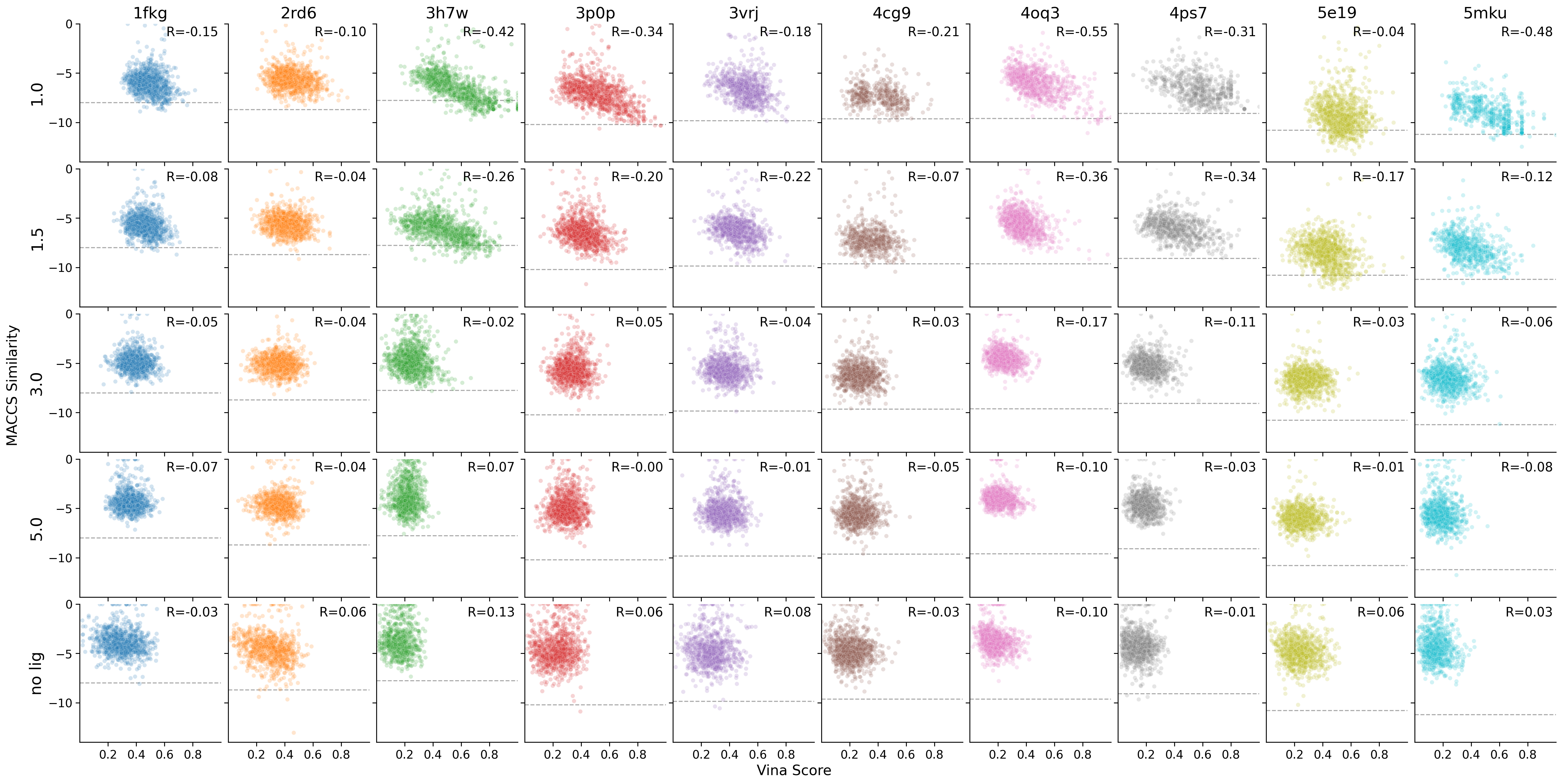}
    \caption{The distribution change in terms of affinity scores and molecular similarity relative to the seed molecules in changing the variable factor}
    \label{fig:cnn_maccs}
\end{figure}

\end{document}